\documentclass[aps,prr,twocolumn,nofootinbib,superscriptaddress,showkeys]{revtex4-1}

\usepackage{color}
\usepackage{graphicx}
\usepackage{dcolumn}
\usepackage{bm}
\usepackage{lipsum}
\usepackage{chngcntr}
\usepackage[mathlines]{lineno}
\usepackage[english]{babel}
\usepackage[latin1]{inputenc}
\usepackage[bottom=2cm,top=2cm, left=1.5cm, right=1.5cm]{geometry}
\usepackage{booktabs}
\usepackage[unicode=true, bookmarks=true, bookmarksnumbered=false, bookmarksopen=false, breaklinks=false, pdfborder={0 0 1}, backref=false, colorlinks=false]{hyperref}

\hypersetup{pdftitle={Terahertz conductivity of heavy-fermion systems from time-resolved spectroscopy}, pdfauthor={Chia-Jung Yang, Shovon Pal, Farzaneh Zamani, Kristin Kliemt, Cornelius Krellner, Oliver Stockert, Hilbert von Loehneysen, Johann Kroha, and Manfred Fiebig}, pdfkeywords={Strong electronic correlations, THz time-domain spectroscopy, Kondo effect, Quantum criticality}}

\begin{document}

\title{Terahertz conductivity of heavy-fermion systems from time-resolved spectroscopy}

\author{Chia-Jung Yang\textsuperscript{\textsection}}
\affiliation{Department of Materials, ETH Z\"{u}rich, 8093 Z\"{u}rich, Switzerland.}

\author{Shovon Pal\textsuperscript{\textsection}}
\email{shovon.pal@mat.ethz.ch}
\affiliation{Department of Materials, ETH Z\"{u}rich, 8093 Z\"{u}rich, Switzerland.}

\author{Farzaneh Zamani}
\affiliation{Physikalisches Institut and Bethe Center for Theoretical Physics, Universit\"{a}t Bonn, 53115 Bonn, Germany.}

\author{Kristin Kliemt}
\affiliation{Physikalisches Institut, Goethe-Universit\"{a}t Frankfurt, 60438 Frankfurt, Germany.}

\author{Cornelius Krellner}
\affiliation{Physikalisches Institut, Goethe-Universit\"{a}t Frankfurt, 60438 Frankfurt, Germany.}

\author{Oliver Stockert}
\affiliation{Max Planck Institute for Chemical Physics of Solids, 01187 Dresden, Germany.}

\author{Hilbert v. L\"{o}hneysen}
\affiliation{Institut f\"{u}r Quantenmaterialien und -technologien and Physikalisches Institut, Karlsruhe Institute of Technology, 76021 Karlsruhe, Germany.}

\author{Johann Kroha}
\email{kroha@th.physik.uni-bonn.de}
\affiliation{Physikalisches Institut and Bethe Center for Theoretical Physics, Universit\"{a}t Bonn, 53115 Bonn, Germany.}

\author{Manfred Fiebig}
\affiliation{Department of Materials, ETH Z\"{u}rich, 8093 Z\"{u}rich, Switzerland.}

\date{August 24, 2020}

\begin{abstract}

The Drude model describes the free-electron conduction in simple metals, governed by the freedom that the mobile electrons have within the material. In strongly correlated systems, however, a significant deviation of the optical conductivity from the simple metallic Drude behavior is observed. Here, we investigate the optical conductivity of the heavy-fermion system CeCu$_{6-x}$Au$_{x}$, using time-resolved, phase-sensitive terahertz spectroscopy. The terahertz electric field creates two types of excitations in heavy-fermion materials: First, the intraband excitations that leave the heavy quasiparticles intact. Second, the resonant interband transitions between the heavy and light parts of the hybridized conduction band that break the Kondo singlet. We find that the Kondo-singlet-breaking interband transitions do not create a Drude peak, while the Kondo-retaining intraband excitations yield the expected Drude response. This makes it possible to separate these two fundamentally different correlated contributions to the optical conductivity.

\end{abstract}

\keywords{Strong electronic correlations, THz time-domain spectroscopy, Kondo effect, Quantum criticality}

\maketitle

\begingroup\renewcommand\thefootnote{\textsection}
\footnotetext{These authors contributed equally to this work}
\endgroup

\section{Introduction} 
The electronic ground state of a strongly correlated system has remained a subject of great interest over decades. Heavy-fermion systems \cite{Stewart1984,Loehneysen2007,Wirth2016,Kirchner2020}, such as the rare-earth intermetallics, are one of the most remarkable manifestations of systems with strong electronic correlations. The localized 4$f$ states of the rare-earth element give rise to local magnetic moments that typically order magnetically at sufficiently low temperatures. In addition, the hybridization between 4$f$ and conduction electrons leads to the Kondo effect \cite{Kondo1964,Hewson1993}, which screens the local moments via the Kondo spin-singlet many-body states. The Kondo effect, thus, drives a part of the 4$f$ spectral weight to the Kondo resonance near the Fermi energy $\varepsilon_{\rm F}$ \cite{Hewson1993}, where it forms a band of lattice-coherent, heavy quasiparticles. These 4$f$ electrons thus become itinerant, leading to the expansion of the Fermi volume, i.e., the $k$-space volume enclosed by the Fermi surface, to accommodate the extra number of indistinguishable 4$f$ electrons in the Fermi sea. The ground state of many heavy-fermion compounds undergoes a second-order quantum phase transition \cite{Gegenwart2007,Friedemann2010,Oliver2011} at the quantum critical point (QCP) \cite{Custers2003}. While in the spin-wave or Hertz-Millis-Moriya scenario \cite{Hertz1976,Moriya1985,Millis1993} of a magnetic quantum phase transition this heavy Fermi liquid undergoes a spin-density wave instability leaving the heavy quasiparticles intact, in some heavy-fermion compounds \cite{Custers2003,Oliver2011,Friedemann2010,Gegenwart2008,Paschen2016,Chen2017} the Kondo screening gets disrupted near the QCP \cite{Coleman2001,Si2001,Hackl2008,Nejati2017}, and the ensuing Kondo breakdown leads to a partial collapse of the Fermi volume \cite{Friedemann2010,Paschen2016,Chen2017}.

\begin{figure*}[t!]
\includegraphics[width=0.7\textwidth]{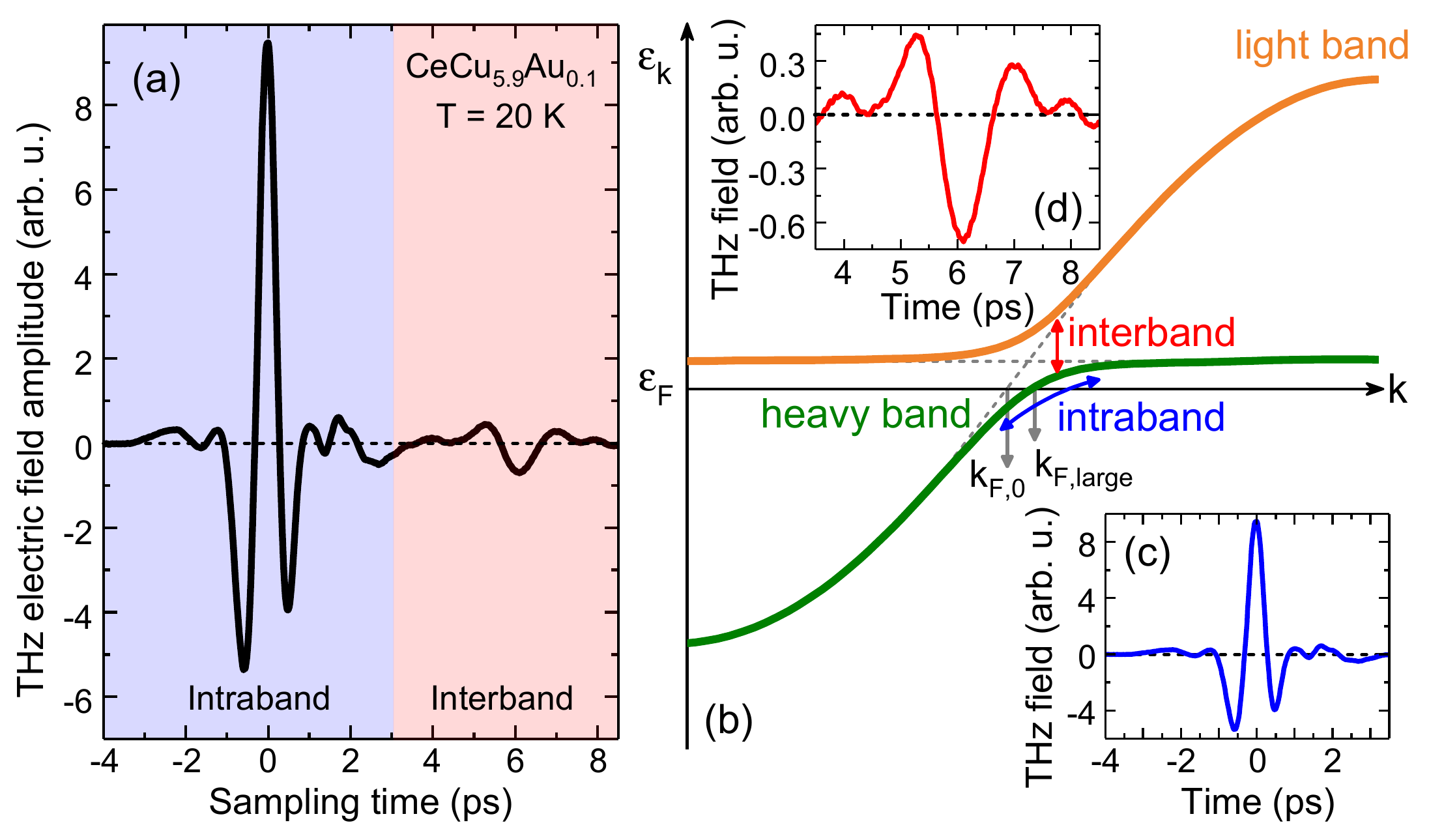}
\caption{(a) The reflected THz signal from the quantum-critical compound, CeCu$_{5.9}$Au$_{0.1}$, at a sample temperature of 20\,K. (b) A schematic of the hybridized band structure showing the intraband excitations and interband transitions as a result of terahertz excitation. Note that intraband excitations are possible due to the presence of impurity or umklapp scattering, like in simple metals, even though the momentum transfer by the THz radiation is negligible. Wavevectors ${\rm k}_{\rm {F,0}}$ and ${\rm k}_{\rm {F,large}}$ correspond to small and large Fermi surface, respectively \cite{Pal2019}. (c) The instantaneous part [blue-shaded region in (a)] resulting from the intraband excitations. (d) The delayed correlated part (red-shaded region in (a)) of the signal featuring interband transitions. The delay time ($\tau^*_{\rm K} = 6$\,ps) corresponds to a Kondo lattice temperature of $T^*_{\rm K} = 8$\,K.}
\label{fig:Figure1}
\end{figure*}

The above-mentioned hybridization between the localized rare-earth 4$f$ electrons and the itinerant conduction electrons control the electronic properties, such as conductivity, in a profound way. At elevated temperatures, the electrons move in a broad conduction band, and infrared optical experiments on the low-energy electrodynamics of heavy-fermion compounds reveal Drude-like behavior in the optical conductivity, indicating light Fermi-liquid nature
\cite{Marabelli1990,Marabelli1989,Scheffler2005,Armitage2009,Basov2011,Seiro2018}. As the temperature is reduced below the lattice Kondo temperature $T_{\rm K}^*$, the Kondo spin scattering of conduction electrons from the rare-earth local moments and the subsequent mixing of nearly localized 4$f$ electron and conduction electron states leads to the formation of a flat band crossing the Fermi level with drastically enhanced effective mass. As a consequence, deviations from the simple-metallic Drude behavior have been observed. These deviations are controlled by two types of phenomena. On one hand, the enhanced quasiparticle effective mass leads to a more inert electronic response to external electric fields and tends to reduce the low-frequency conductivity. On the other hand, the Kondo correlation-enhanced density of states near the Fermi level tends to increase the conductivity. In addition, the enhanced spin-scattering rate at temperatures around $T_{\rm K}^*$, reflecting the crossover from the light to the heavy Fermi liquid, is expected to broaden the Drude peak in the optical conductivity as a function of frequency. The interplay of these latter many-body effects with the former effective-mass, i.e., band-structure, effect can lead to complex temperature dependence of the optical conductivity \cite{Scheffler2005,Scheffler2010}. Up to now, it has been difficult to disentangle the correlation-induced many-body dynamics from single-particle bandstructure effect in physical observations. This is partly because the Kondo resonance and the Drude behavior in the low-frequency conductivity response both have a Lorentzian line shape. Finding a proper way to accomplish this separation is thus of fundamental importance, in particular, to obtain a deeper understanding of the complex behavior of the conductivity which has been apparent in optical-conductivity experiments on heavy-fermion compounds for decades.

Recently, time-resolved terahertz (THz) spectroscopy has proven to be a powerful tool to coherently probe collective excitations in solids, in particular the correlation dynamics in heavy-fermion systems. We have shown that upon excitation with low-energy THz radiation, part of the correlated Kondo state can be extinguished and resurges back with a distinct temporal separation from the instantaneous conduction-electron response \cite{Wetli2018}. In order to identify this delayed response within the time traces, the following conditions must be fulfilled: (1) The energy scale of the Kondo correlated states, i.e.\ the Kondo lattice temperature $T^*_{\rm K}$, should be well separated from the single-particle binding energy of the 4$f$ electrons in the heavy-fermion system (Kondo regime). (2) The incident pulse duration should be short compared to the delay time, given by the Kondo quasiparticle lifetime $\tau_{\rm K}^*= \hbar /k_{\rm B}T^*_{\rm K}$, where $\hbar$ is the reduced Planck constant and $k_{\rm B}$ the Boltzmann constant. The delayed response bears distinct information on the Kondo correlation dynamics. It entails a rather direct measure of $T^*_{\rm K}$ and of the quasiparticle spectral weight of the heavy-fermion system within a single experiment \cite{Wetli2018,Pal2019}. In short, the THz electric field creates (i) the intraband excitations, which leave the heavy quasiparticles intact and lead to an instantaneous response, and (ii) the resonant interband transitions, which break the Kondo singlets and lead to a time-delayed response. Taking advantage of this temporal separation, we can now discern the optical conductivity obtained from the instantaneous and the delayed, correlated responses. Moreover, as our technique is phase-sensitive, one can avoid invoking the Kramers-Kronig analysis while extracting the real and imaginary parts of the optical conductivity. This leads to two questions: First, does the optical conductivity obtained in our current approach from the instantaneous response reconcile with the earlier observations from infrared experiments? Second, what is the behavior of the dynamical conductivity obtained from the delayed, correlated response?

In this Report, we address these questions by unravelling the temperature dependence of the dynamical conductivity obtained from the instantaneous response and the correlated delayed response at THz frequencies. We use the canonical CeCu$_{6-x}$Au$_{x}$ system \cite{Loehneysen1994,Schroeder2000}, with ${x} = 0$ as the heavy-fermion compound, ${x} = 0.1$ as the quantum-critical compound and ${x} = 1$ as the antiferromagnetic compound. It is important to note that CeCu$_5$Au is a stoichiometric compound where exactly one of the five inequivalent Cu sites is completely occupied by Au \cite{Ruck1993}. A Pt mirror is used as the reference sample for all our measurements. We show that the dynamical conductivity obtained from the instantaneous response of CeCu$_6$ agrees well with previous observations \cite{Marabelli1990,Marabelli1989}, i.e., it shows a metallic Drude-like low-frequency behavior at high temperatures that develops into a heavy-Fermi-liquid Drude response at $T \le T^*_{\text{K}}$. In contrast, the dynamical conductivity obtained from the delayed response does not show any Drude-like free-electron behavior. Such a non-trivial deviation corroborates that the delayed response is a unique signature of the correlated electronic states.

\section{Experimental} 
The CeCu$_{6-x}$Au$_{x}$ samples used in this work are cut from single crystals of orthorhombic crystal structure, and faces perpendicular to the crystallographic $c$-axis were polished using colloidal silica. In order to prevent oxidation, all samples were stored in an inert nitrogen atmosphere. The sample surface is always freshly polished before measurements, and care is taken to ensure that the surface damages due to polishing are within the submicron range (i.e., $\ll \lambda_{\text{THz}}$). The samples are then mounted in a temperature-controlled Janis SVT-400 helium reservoir cryostat. The experiments are performed in a reflection geometry where linearly polarized THz radiation with a spectral range of 0.1 - 3\,THz is used at an incident angle of 45$^{\circ}$. The THz electric field is oriented perpendicular to the crystallographic $a$-axis.

Single-cycle terahertz pulses are generated by optical rectification in a 0.5\,mm ZnTe(110)-oriented single crystal, using 90\% of a Ti:Sapphire laser output (wavelength 800\,nm, pulse duration 120\,fs, pulse repetition rate 1\,kHz, pulse energy 2\,mJ). The remaining 10\% of the fundamental pulse is used as a probe (or gating) pulse for the free-space electrooptic sampling of the reflected THz wave. The THz and the gating beams are collinearly focused onto a ZnTe(110)-oriented detection crystal. The THz-induced ellipticity of the probe light is measured using a quarter-wave plate, a Wollaston polarizer and a balanced photodiode. The signal from the photodiode is then analyzed with a lock-in amplifier. In order to increase the accessible time delay between the THz and the probe pulses, Fabry-P\'erot resonances from the faces of the 0.5\,mm ZnTe(110)-oriented crystal are suppressed by extending the detection crystal with a 2\,mm THz-inactive ZnTe(100)-oriented crystal that is optically bonded to the back of the detection crystal. 

\begin{figure}[t!]
\includegraphics[width=\columnwidth]{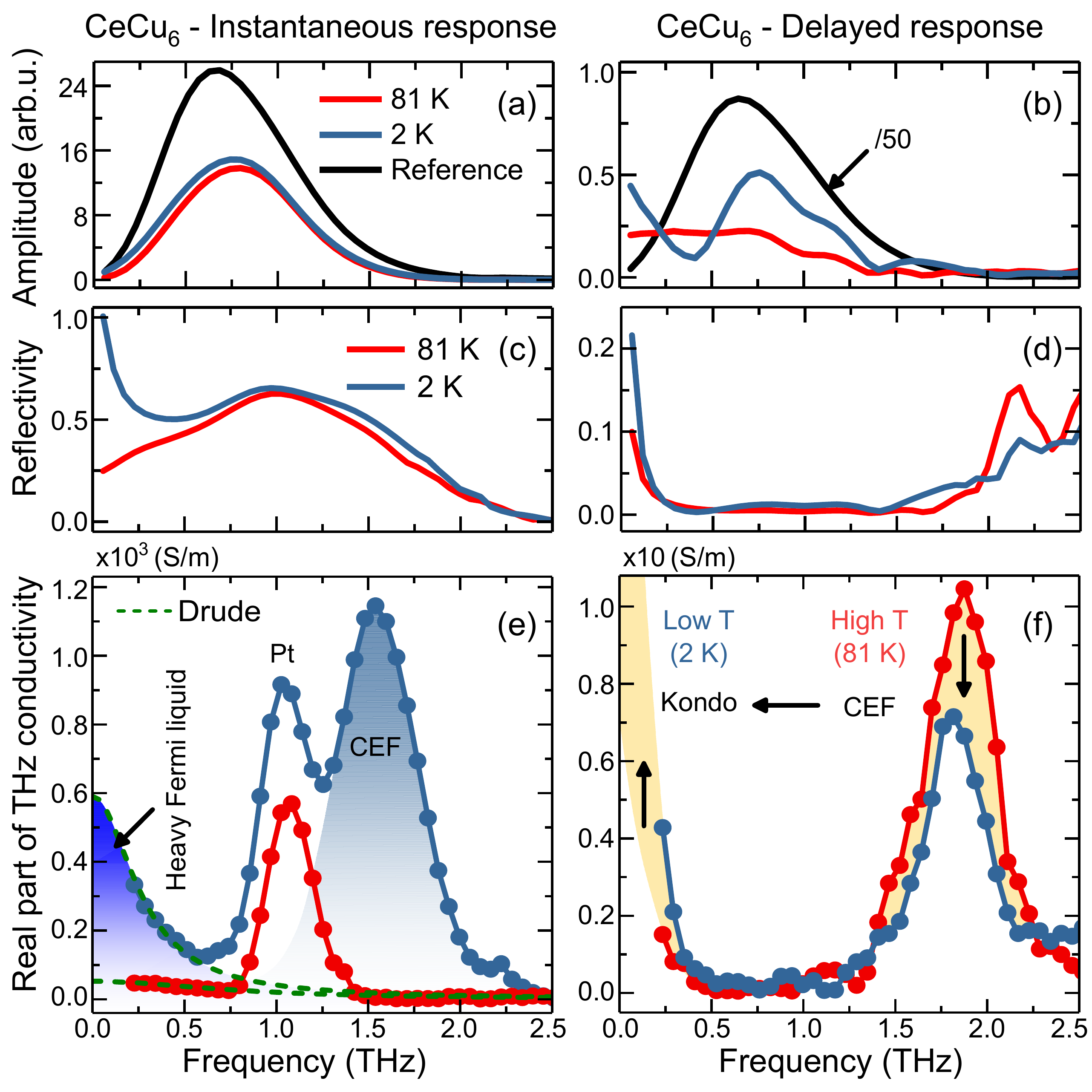}
\caption{(a),\,(b) The spectra obtained by Fourier transforming the
  instantaneous and delayed THz responses of the heavy-fermion compound
  (CeCu$_6$) at sample temperatures of 81\,K and 2\,K. The reference spectrum
  is the signal from the Pt mirror. (c),\,(d) The reflectivity curves at 81\,K and 2\,K, corresponding to the instantaneous and delayed THz responses. (e) The real part of the conductivity obtained from the instantaneous response shows a heavy-Fermi-liquid behavior (green dashed curve) in the low frequency region at 2\,K and a peak corresponding to the CEF resonance at higher frequencies. (f) The real part of the conductivity obtained from the correlated delayed response shows the low-frequency Kondo resonance and the high-frequency CEF resonance. A smooth transfer of the correlated spectral weight from the CEF resonance at high temperatures to the Kondo resonance at low temperatures is highlighted by the arrows in the yellow-shaded regions.}
\label{fig:Figure2}
\end{figure}

\section{Results and Discussion} 
We first identify the trivial THz reflexes generated by the optical components in order not to confuse them with the THz signal generated from the CeCu$_{6-x}$Au$_{x}$ samples. A measurement of all these THz reflexes is provided elsewhere \cite{Wetli2018}. The earliest of these artifacts appear at a delay time of 10\,ps, outside the range used in our analysis for the optical conductivity. As an example, we show the time trace of the THz electric field reflected from the CeCu$_{5.9}$Au$_{0.1}$ sample at 20\,K in Fig.~\ref{fig:Figure1}(a). The complete signal can be divided into two parts: (i) the instantaneous response,
shown in Fig.~\ref{fig:Figure1}(c), from $t = -4$\,ps to $t = +3.5$\,ps [blue-shaded region in Fig.~\ref{fig:Figure1}(a)], and (ii) the delayed response, shown in Fig.~\ref{fig:Figure1}(d) from $t = +3.5$\,ps to $t = +8.5$\,ps [red-shaded region in Fig.~\ref{fig:Figure1}(a)]. As mentioned previously, these two temporally separated responses stem from two excitation processes induced by the THz radiation, the intraband excitations and the interband transitions. The schematic in Fig.~\ref{fig:Figure1}(b) shows the hybridized band, elucidating the two processes. The instantaneous response originates from the stimulated single-particle response of quasiparticles within the conduction band, in other words, intraband excitations. These intraband scattering processes leave the heavy quasiparticles intact and are expected to show a Fermi-liquid Drude response in the optical conductivity. The delayed response, on the other hand, originates from the interband transitions between the hybridized heavy and light parts of the conduction band that restores the Kondo state after the stimulated breaking (by the THz pulse) of the Kondo singlets. In CeCu$_{6-x}$Au$_{x}$, the delayed response appears around 6\,ps and agrees well with the Kondo quasiparticle lifetime $\tau_{\rm K}^*$ \cite{Wetli2018}. Since this response comes solely from the Kondo correlation effect, we would not expect to observe a Drude response but rather a low-frequency Kondo resonance with a Lorentzian lineshape. The crystal-electric-field (CEF) resonances play a special role in heavy-fermion compounds since at high temperatures the CEF-induced high-energy scale governs the spectral weight near the Fermi level \cite{Pal2019}. At low temperature, they appear as a set of resonances with narrow spectral width \cite{Reinert2001,Ehm2007} and, hence, long coherence time. Their signature is therefore primarily found in the delayed pulse. With increasing temperature, the resonances merge into a single peak of increased spectral width and, hence, short coherence time so that their weight shifts from the delayed pulse into the wiggles of the instantaneous pulse \cite{Pal2019}.

\begin{figure*}[t!]
\includegraphics[width=\textwidth]{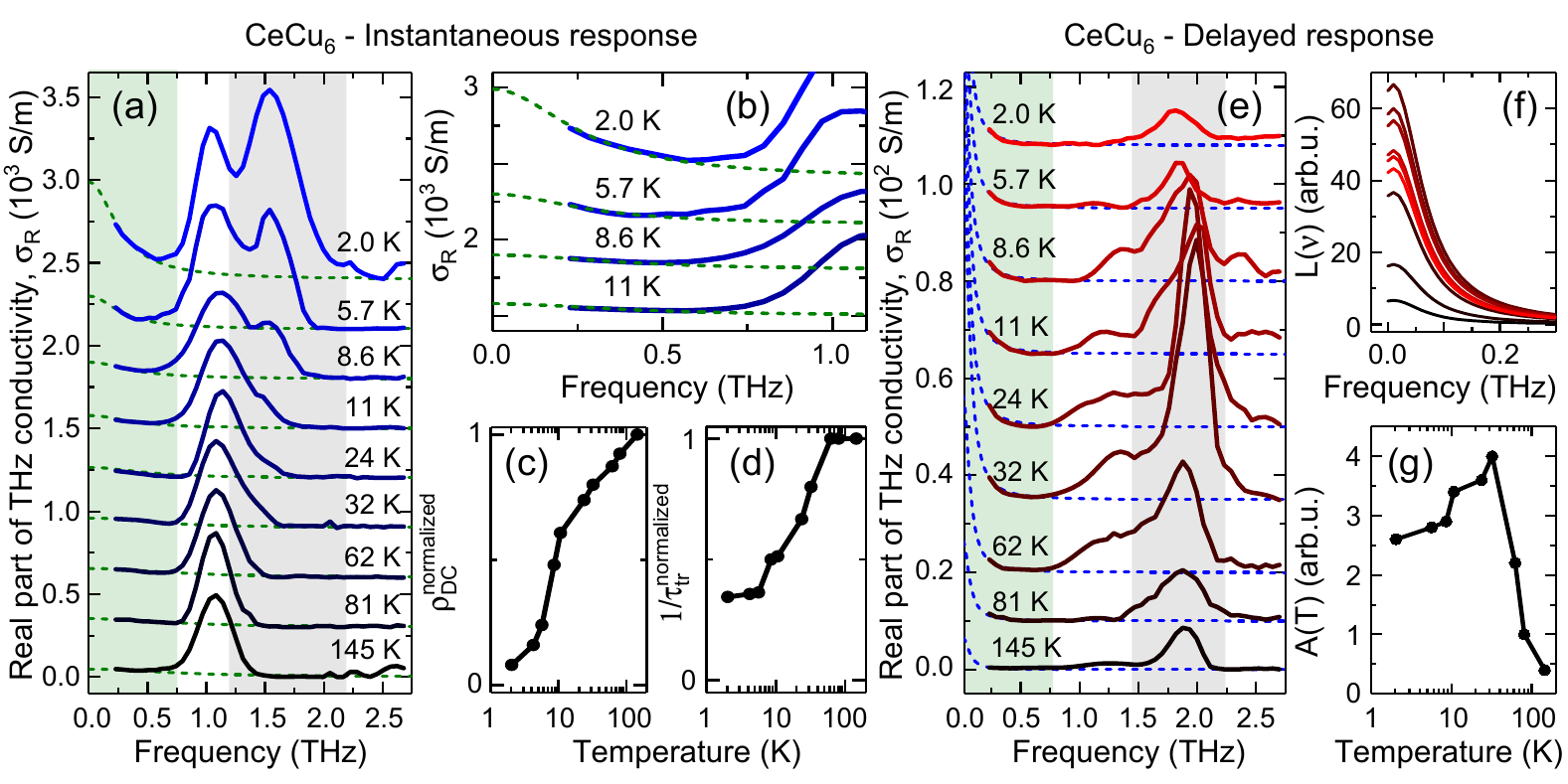}
\caption{(a) The real part of the optical conductivity of CeCu$_6$ obtained from the instantaneous response as a function of sample temperature. The green-shaded region indicates the low-frequency region where the Drude-like free-electron response at high temperatures deviates to a heavy-Fermi-liquid behavior at low temperatures. The gray-shaded region indicates the region where the first CEF resonance starts appearing as the temperature is reduced below 10\,K. The green-dashed lines are the Drude fitting curves. (b) The low-frequency region in (a) for a few selected temperatures. (c),\,(d) The change in DC resistivity and scattering rate, obtained from low-frequency Drude fit, normalized to the value at 145\,K. (e) The real part of the optical conductivity of CeCu$_6$ obtained from the delayed correlated response as a function of sample temperature. Being a pure correlated response, a single peak Lorentz function (blue-dashed curves) with a fixed linewidth of 0.12\,THz is used to fit the low-frequency region (green-shaded region). The CEF resonance (in gray-shaded region) shows the characteristic temperature dependence where the spectral weight from the CEF resonance is transferred to the Kondo resonance on reducing the temperature. (f) The temperature-dependent Lorentz functions used to reproduce the low-frequency conductivity response [all curves are colored as in (e)]. (g) The temperature-dependent amplitude of the single peak Lorentz functions. Note that these reproduce the heavy-fermion Kondo-spectral-weight behavior of Ref.~\cite{Wetli2018}.}
\label{fig:Figure3}
\end{figure*}

\begin{figure*}[t!]
\includegraphics[width=\textwidth]{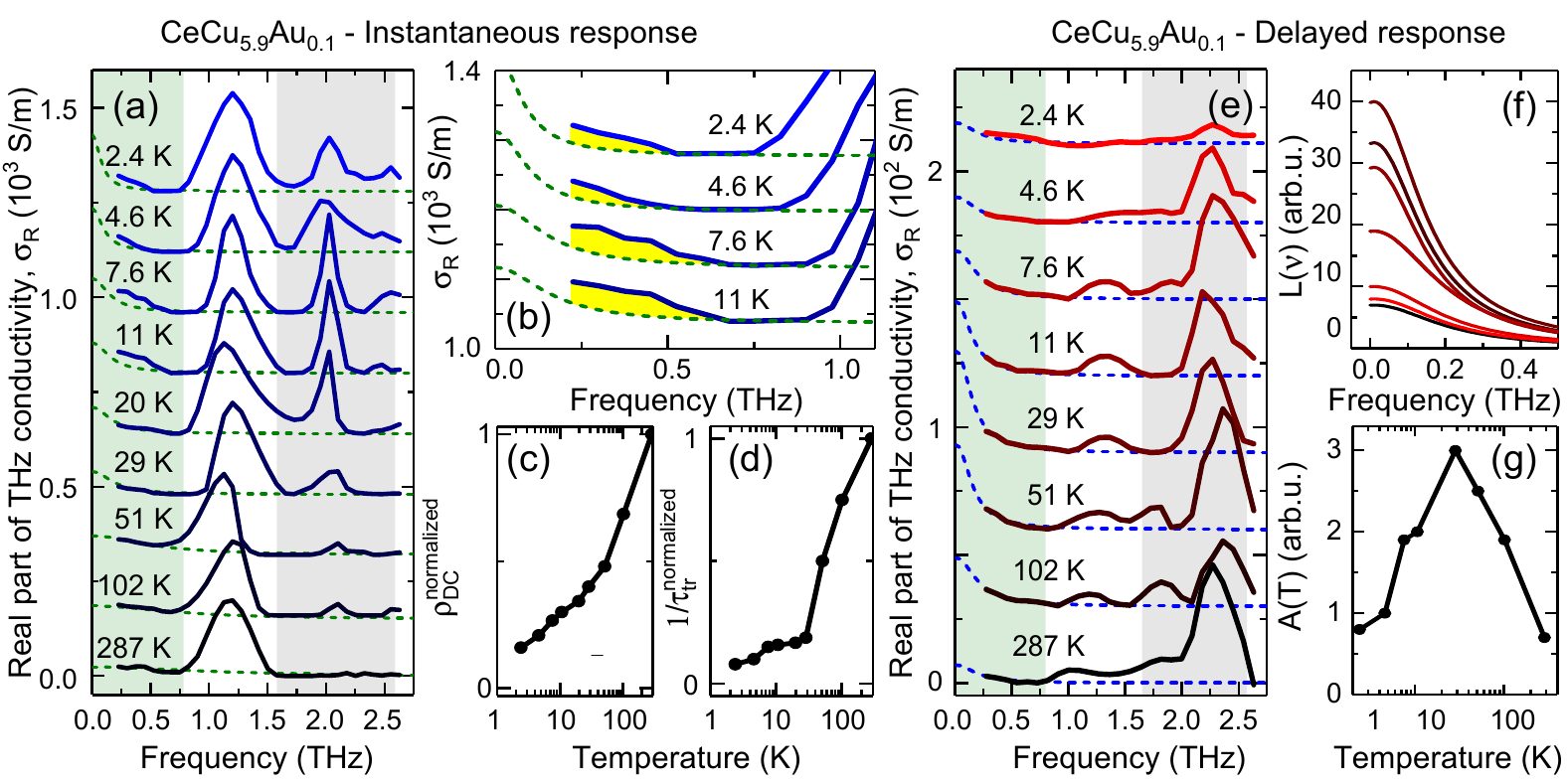}
\caption{(a) The real part of the optical conductivity of CeCu$_{5.9}$Au$_{0.1}$ obtained from the instantaneous response as a function of sample temperature. The green-shaded region indicates the low-frequency region where the Drude-like free-electron response at high temperatures deviates to a non-Fermi-liquid behavior at low temperatures. The gray-shaded region indicates the region where the first CEF resonance starts appearing as the temperature is reduced below 10\,K. The green-dashed lines are the Drude fitting curves. (b) The low-frequency region in (a) for a few selected low temperatures, where the yellow-shaded regions mark the deviations from Drude-like free-electron to non-Fermi-liquid behavior. (c),\,(d) The change in DC resistivity and scattering rate, obtained from low-frequency Drude fit, normalized to the value at 287\,K. (e) The real part of the optical conductivity of CeCu$_{5.9}$Au$_{0.1}$ obtained from the delayed correlated response as a function of sample temperature. Being a pure correlated response, a single peak Lorentz function (blue-dashed curves) with a fixed linewidth of 0.3\,THz is used to fit the low-frequency region (green-shaded region). (f) The temperature-dependent Lorentz functions used to reproduce the low-frequency conductivity response [all curves are colored as in (e)]. (g) The temperature-dependent amplitude of the single peak Lorentz functions. Note that these reproduce the quantum-critical Kondo-spectral-weight behavior of Ref.~\cite{Wetli2018}.}
\label{fig:Figure4}
\end{figure*}

To obtain the dynamical conductivity from the two response windows, the THz
time traces are first Fourier-transformed (FT) to obtain the complex THz
spectra. Figures \ref{fig:Figure2}(a) and \ref{fig:Figure2}(b) show the FT
spectral amplitude of the instantaneous and the delayed responses for CeCu$_6$
at two different temperatures (81\,K, red; and 2\,K, blue), respectively. The
complex reflectivities are obtained by dividing the respective spectra by the
Pt mirror reference spectrum [black curve in Fig.~\ref{fig:Figure2}(a)]. The
corresponding reflectivities are shown in Figs.~\ref{fig:Figure2}(c) and \ref{fig:Figure2}(d). The reflectivity of CeCu$_6$ at 2\,K clearly shows strong deviation from the one at 81\,K, particularly at low frequencies ($\nu \le 0.75$\,THz) and at higher frequencies ($1.2$\,THz $\le \nu \le 2.2$\,THz). Figures \ref{fig:Figure2}(e) and \ref{fig:Figure2}(f) show the real part of the THz conductivity evaluated from the complex reflectivity at 81\,K and 2\,K, for the instantaneous and delayed responses, respectively. The green-dashed curves in Fig.~\ref{fig:Figure2}(e) show the low-frequency Drude fit according to the relation \cite{Basov2011,Scheffler2005,Drude1900,Dressel2006} $\sigma_{\rm R} = \sigma_{\rm DC}/(1+\omega^2\tau_{\rm tr}^2)$, where $\sigma_{\rm R}$ is the real part of the conductivity, $\sigma_{\rm DC}$ is the DC conductivity, $\omega = 2\pi \nu$ is the angular frequency and $1/\tau_{\rm tr}$ is the transport relaxation rate. The Drude fit is carried out only for the range from 0.2 to 0.75\,THz, with $\sigma_{\rm DC}$ and $\tau_{\rm tr}$ being the only fitting parameters. The fit was pinned to the data at the highest frequency, ignoring all spectral features between 1.0 to 1.8\,THz, to illustrate the physically meaningful situation that develops at low frequencies, i.e., $\nu \le 0.75$\,THz. The DC value of the optical conductivity is obtained from the extrapolation of the finite frequency results to zero frequency \cite{Marabelli1990,Marabelli1989,Prochaska2020}. Because of the large overall systematic error in the absolute value of the DC conductivity and the scattering rate, we revert to relative values for their temperature dependence by taking the ratio with respect to the high temperature values (here, with $T = 145$\,K for CeCu$_6$) in the ensuing discussions. The temperature-dependent imaginary part of the conductivities for all samples are shown in Appendix C.

\begin{figure}[t!]
\includegraphics[width=0.8\columnwidth]{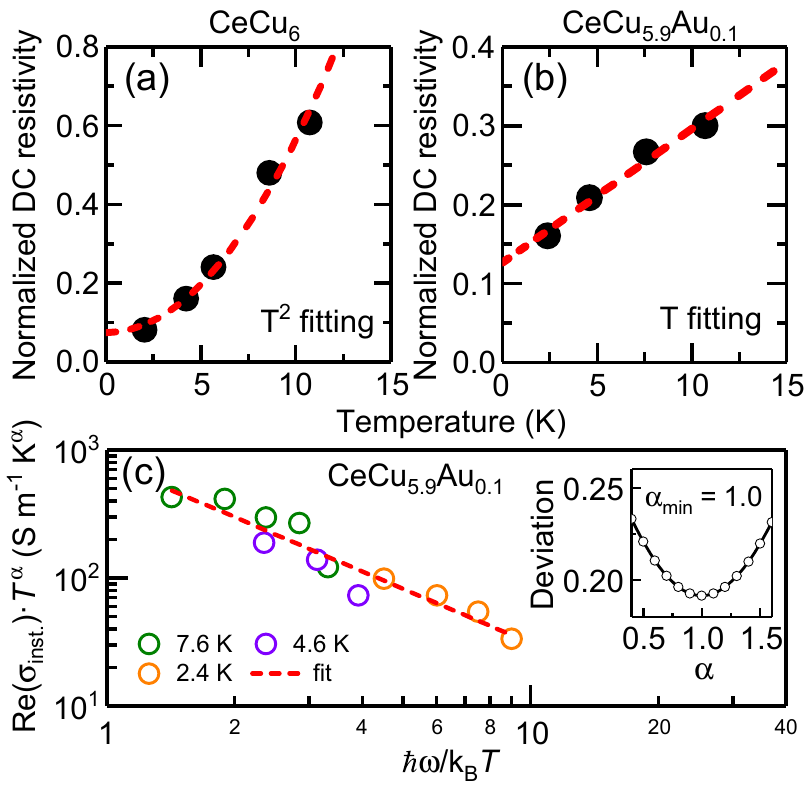}
\caption{The low temperature normalized DC resistivity for (a) heavy-fermion, CeCu$_6$ and (b) quantum-critical, CeCu$_{5.9}$Au$_{0.1}$ compounds. The red-dashed curve represents the corresponding low-temperature $T^2$ and $T$ fitting of the resistivity data, in accordance with Fermi-liquid and non-Fermi-liquid behavior, respectively. (c) $\omega/T$ scaling plot of the optical conductivity (real part) for CeCu$_{5.9}$Au$_{0.1}$ taken from the instantaneous response, showing approximate scaling collapse in the range $1\lesssim \hbar\omega/k_{\rm B}T\lesssim 9$. The exponent of the $\hbar \omega/k_{\rm B}T$ dependence is 1.3. The inset shows the standard deviation of the scaling fit in dependence on the exponent $\alpha$.}
\label{fig:Figure5}
\end{figure}

In Fig.~\ref{fig:Figure2}(e), we find that the instantaneous low-frequency conductivity at 81\,K qualitatively follows the metallic Drude response and, as the temperature is reduced to 2\,K, the conductivity deviates from the simple-metallic Drude nature to a heavy-Fermi-liquid Drude behavior, which is characterized by a decreased electronic relaxation rate [see Fig.~\ref{fig:Figure3}(d)] and an enhanced effective mass and density of states near the Fermi level. The gradual appearance of the peak around 1.5\,THz in the optical conductivity below 40\,K [see also Fig.~\ref{fig:Figure3}(a)] indicates the formation of a narrow quasiparticle band in the vicinity of the Fermi level, in good agreement with de Haas-van Alphen measurements \cite{Reinders1986}. Thus the changes observed in the low-temperature data result from the interaction of the delocalized 4$f$ electrons, i.e., the formation of heavy quasiparticles. The conductivity peak at 1\,THz is the free-carrier response in the Pt film used as a reference. The Pt film is no perfect mirror and therefore has a distinct frequency response compared to the incident beam (or to the original THz beam that is incident on the sample). In the frequency-domain analysis, we obtain this as a conductivity peak. We carried out the conductivity analysis of the Pt mirror taking the bare THz pulse as a reference, see Fig.~\ref{fig:reference} in Appendix B. This analysis clearly reveals that we have a distinct peak at 1 THz, which is temperature-independent with none of the additional features that we obtain from the sample. Hence, we ignore it in the rest of our discussion. 

In contrast, the optical conductivity obtained from the delayed pulse around 6\,ps shows a distinct response \cite{Wetli2018,Pal2019} arising from the break-up and restoration of the correlated Kondo state, see Fig.~\ref{fig:Figure2}(f). Within the spectral range, two distinct features signaling the heavy-fermion state are observed. First, we have a quasi-Lorentzian peak at low frequencies, $\nu \le 0.75$\,THz, that is fundamentally different from the Drude response [green-dashed curve in Fig.~\ref{fig:Figure2}(e)]. The spectral weight of this low-frequency resonance peak increases as the temperature is reduced due to the increasing Kondo weight. The width of this resonance peak resembles the Kondo energy scale, which remains constant though the spectral weight reduces towards low temperatures, in agreement with our previous findings \cite{Wetli2018, Pal2019}. The second feature is the broad peak centered at 1.75\,THz that corresponds to the first CEF-excited state in the CeCu$_6$ system \cite{Stroka1993,Goremychkin1993,Witte2007}. The spectral weight of this peak reduces as the temperature is reduced. Note that the occupied spectral weight of the CEF resonance at higher temperatures is smoothly transferred to the Kondo resonance at low temperatures [indicated schematically by the yellow-shaded regions in Fig.~\ref{fig:Figure2}(f)] as the CEF occupation is frozen out \cite{Reinert2001,Ehm2007}. This observation agrees with our previous results, drawn from the time-domain, on the smooth crossover from a high-temperature to a low-temperature Kondo scale \cite{Pal2019}. Also note that the low-temperature transport relaxation rate extracted from the Drude fit in Fig.~\ref{fig:Figure2}(e) as $1/\tau_{\rm tr}\approx 0.25$\,THz  (corresponding to 2\,K) is significantly larger than the Kondo energy scale (equivalent to $T_{\rm K}^*\approx 8$\,K) and of the width of the low-frequency peak in the delayed response of Fig.~\ref{fig:Figure2}(f). This indicates that the low-frequency instantaneous response is governed by intraband transport scattering processes and concatenated Drude behavior [Fig.~\ref{fig:Figure2}(e)], while the zero-frequency peak in the delayed response [Fig.~\ref{fig:Figure2}(f)] is a signature of the Kondo resonance alone.

A complete temperature evolution of the THz conductivity obtained from the
instantaneous signal of the heavy-fermion compound, CeCu$_6$, is shown in
Figs.~\ref{fig:Figure3}(a) and \ref{fig:Figure3}(b). From the low-frequency
Drude fitting, mentioned above, we could retrieve the temperature dependence
of the DC conductivity and hence the resistivity as well as the scattering
rate, shown in Figs.~\ref{fig:Figure3}(c) and \ref{fig:Figure3}(d),
respectively. We see that at very high temperatures, the real part of the THz
conductivity is dominated by the Drude-like free-electron response, as
mentioned earlier. This behavior, however, develops slowly into a
heavy-Fermi-liquid Drude response as the temperature is lowered below 10\,K,
characterized by the reduced scattering rate and increased effective mass,
with a $T^2$ dependence of resistivity featuring the Fermi-liquid intraband
quasiparticle-quasiparticle scattering [see red-dashed line in
  Fig.~\ref{fig:Figure5}(a)]. The gray-shaded region in
Fig.~\ref{fig:Figure3}(a) shows the evolution of the first CEF resonance. 
Similar low-temperature behavior was observed earlier for CeCu$_6$
from infrared measurements \cite{Marabelli1990}. This certifies a strong $f$ character of the CEF bands at the Fermi level stemming from the hybridization between the conduction electrons and the 4$f$ electrons. 

The temperature evolution of the conductivity obtained from the delayed response [Fig.~\ref{fig:Figure3}(e)] is striking. The low-frequency behavior cannot be reproduced using the Drude-like response of width $1/\tau_{\rm tr}$ at any temperature. However, we can fit a single peak Lorentz function of smaller width $\Gamma$ [blue-dashed lines in Fig.~\ref{fig:Figure3}(e)]. The function is defined as $L(\nu) = A(T)\Gamma/[(\nu-\nu_0)^2+(\Gamma/2)^2]$, where $A(T)$ is the temperature-dependent spectral weight, $\nu_0$ represents the low-frequency Kondo resonance and $\Gamma$ should correspond to the Kondo energy scale, see discussion above. The temperature dependence of the fitted Lorentz functions is separately plotted in Fig.~\ref{fig:Figure3}(f). The temperature dependence of $A$, plotted in Fig.~\ref{fig:Figure3}(g), reproduces qualitatively the temperature dependence of the Kondo weight in CeCu$_6$, obtained by integrating the spectral weight of the delayed response \cite{Wetli2018,Pal2019}. At high temperatures there is only one peak observed at around 1.75\,THz which corresponds to the first CEF resonance. As the temperature is reduced below 300\,K the spectral weight of this peak first increases down to 40\,K, revealing the high-temperature Kondo scale \cite{Pal2019}. On decreasing the temperature further, the spectral weight of the peak reduces while the spectral weight of the low-frequency Kondo resonance increases [see Fig.~\ref{fig:Figure2}(f) for clarity]. This indicates that there is a smooth transfer of the spectral weight from the high-temperature CEF resonance to the low-temperature Kondo resonance, as mentioned before. Further, we observe that roughly $A(T \to 0) \approx 0.7 \times A(T = 35$\,K), in very good agreement with Ref.~\cite{Wetli2018}.

In the case of the quantum-critical sample, CeCu$_{5.9}$Au$_{0.1}$, the
optical conductivity obtained from the instantaneous response [see
  Fig.~\ref{fig:Figure4}(a) and ~\ref{fig:Figure4}(b)] shows a very similar
high-temperature Drude behavior as observed in the case of the heavy-fermion
compound, CeCu$_6$. However, for $T < 10$\,K the low-frequency part of the
conductivity shows deviation from the Drude behavior that presumably reflects
the onset of the non-Fermi-liquid behavior \cite{Prochaska2020,Berthod2013}
near the quantum-critical point [see the yellow-shaded region in
  Fig.~\ref{fig:Figure4}(b)]. From the low-frequency Drude fitting, the
temperature dependence of the DC conductivity as well as the scattering rate
are obtained as explained above and are shown in Figs.~\ref{fig:Figure4}(c)
and ~\ref{fig:Figure4}(d), respectively. The resistivity behavior is
distinctly different from CuCu$_6$ for $T < 10$\,K. While CeCu$_6$ has a $T^2$
resistivity dependence [see Fig.~\ref{fig:Figure5}(a)], CeCu$_{5.9}$Au$_{0.1}$
shows a resistivity compatible with linear-in-$T$ dependence for $T< 10\,K$
[see Fig.~\ref{fig:Figure5}(b)], in line with the linear DC resistivity found
below 1\,K \cite{Loehneysen1994} This difference can be associated with the
fact that, while CeCu$_6$ is a heavy-Fermi-liquid compound at these
temperatures, CeCu$_{5.9}$Au$_{0.1}$ is quantum critical. The temperature dependence of the conductivity from the delayed response [see Fig.~\ref{fig:Figure4}(e)], is once again striking. A single-peak Lorentz function is used, as before, to model the low-frequency conductivity behavior, see Fig.~\ref{fig:Figure4}(f). We find that the spectral weight of the low-frequency Kondo resonance increases as we lower the temperature, reaches a maximum and then smoothly reduces as we lower the temperature below $T_{\rm K}^*$. This corroborates that the sample is entering into the quantum-critical regime where the Kondo quasiparticle weight almost completely dissappears, again in agreement with Ref.~\cite{Wetli2018}, see Fig.~\ref{fig:Figure4}(g). Further, the high-frequency CEF resonance also reduces its spectral weight as we lower the temperature, quite similar to what is observed in CeCu$_6$. These results reproduce our previous work \cite{Wetli2018,Pal2019}, where our analysis was carried out purely from the responses in the time domain. Thus, we consistently agree with the physics of the Kondo-correlated part from the delayed, echo-like response in both time and the frequency domain. The spectral weight in the time domain corresponds to the optical conductivity in the frequency domain. Yet, in addition to the earlier results, a proper identification of the pure correlated response in the time-domain enables us to discern (i) the Kondo resonance from the Drude behavior and (ii) the effect of CEF states on the high-energy Kondo scale.

Near a Kondo-breakdown QCP, when the low-energy Fermi liquid scale $T^{*}$
vanishes, dynamical response quantities should obey $\omega/T$ scaling of the
form $\sigma(\omega,T)\propto T^{-\alpha}\,f(\hbar\omega/k_{\rm B}T)$ with a
universal scaling function $f(x)$. This was theoretically expected
\cite{Zhu2004,Komijani2019} and experimentally found not only in
CeCu$_{5.9}$Au$_{0.1}$ for the dynamic magnetic response \cite{Schroeder2000},
but also more recently in YbRh$_2$Si$_2$ for the optical conductivity
\cite{Prochaska2020}. Note that $T^*$ is to be distinguished from the Kondo
lattice temperature $T_{\rm K}^*$ characterizing the onset of Kondo
spin-screening and heavy quasiparticle formation. For our scaling analysis, we
use the instantaneous (i.e., intraband) response data $\sigma_{\text{inst}}$
[in Fig.~\ref{fig:Figure4}(b) for $T < T^*_{\rm K}$], since here we are
interested in the low-energy excitations in the limit $\omega\to 0$. In
Fig.~\ref{fig:Figure5}(c) we present a scaling plot of the real part,
Re$[\sigma_{\rm {inst.}}(\omega)]\,T^{\alpha}$, as a function of
$\hbar\omega/(k_{\rm B}T)$. We find that, within experimental accuracy, all
the isotherms below $T_{\rm K}^*\approx 8$\,K and frequencies below 1\,THz
collapse onto a single curve for a critical exponent of $\alpha \approx$ 1.0,
demonstrating $\omega/T$ scaling in the region of $T$-linear resistivity for
the quantum-critical compound. The slope of the red, dashed line fitted to the
data in Fig.~\ref{fig:Figure5}(c) is 1.3. See Appendix D for details of the
$\omega/T$ scaling analysis. Our values of the scaling exponents are
consistent, within experimental accuracy, with the values extracted in
Ref.~\cite{Prochaska2020} for YbRh$_2$Si$_2$. The $\omega/T$ scaling points to
the fact that the Kondo-breakdown scenario seems to be valid in the
CeCu$_{6-x}$Au$_{x}$ system, in agreement with our earlier findings for the
Kondo spectral-weight breakdown \cite{Wetli2018,Pal2019}.

\begin{figure}[t!]
\includegraphics[width=\columnwidth]{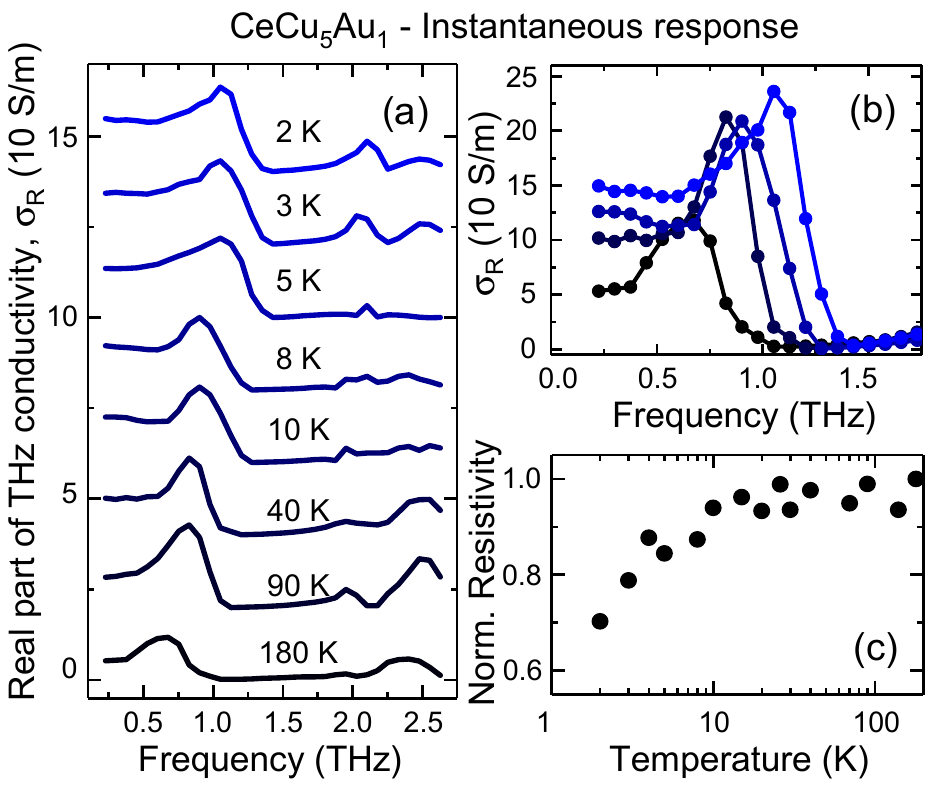}
\caption{(a) The real part of the optical conductivity of CeCu$_{5}$Au$_{1}$ obtained from the instantaneous response as a function of sample temperature. (b) The low-frequency conductivity response for a few selected temperatures (all curves are colored as in (a)). (c) The normalized frequency-averaged resistivity obtained from (a) as a function of temperature, normalized with respect to the value at 180\,K.}
\label{fig:Figure6}
\end{figure}

Figure.~\ref{fig:Figure6}(a) shows the temperature evolution of the optical
conductivity for the antiferromagnetic compound, CeCu$_{5}$Au$_{1}$, obtained
from the instantaneous response. Remarkably, the antiferromagnetic compound
does not show a delayed correlated response at any temperatures
\cite{Wetli2018}. Apart from the conductivity peak at 1\,THz that, as
mentioned before, is present in all samples, there seems to be a convoluted
underlying broad low-frequency feature that blue-shifts as we lower the
temperature [see Fig.~\ref{fig:Figure6}(b) for clarity]. For higher
temperatures, the intensity from the Pt peak appears to reduce, as observed in
the bare Pt conductivity response (see Fig.~\ref{fig:reference} in Appendix
B). Such an underlying feature makes it difficult to perform a conclusive
Drude fit of the experimental data. We thus speculate about the origin of our
observations as follows: It is known from previous temperature-dependent
nuclear magnetic resonance studies that the nuclear spin-lattice relaxation
mechanism in CeCu$_5$Au is rather anisotropic. For crystallographic
orientations orthogonal to the $c$ axis (as in our current configuration), nuclear spin-spin coupling prevails, mediated by conduction electrons \cite{Kerscher2001}. This implies that temperature-dependent fluctuations in the spin relaxation time can significantly modify the electronic scattering times, thus leading to the observed temperature-dependent complex low-frequency conductivity response. Taking a closer look at the high-frequency response of the conductivity at 2\,THz, it can be argued that a resonance structure appears as the temperature is lowered, possibly indicating the emergence of CEF correlated states. This feature is caused by incoherent Kondo signatures present at all temperatures in our earlier time-domain analysis (see supplement of Ref.~\cite{Wetli2018}). Such incoherent Kondo signatures have also been observed in previous transport and thermodynamic measurements \cite{Loehneysen1998}, despite the dominance of RKKY interaction over the Kondo interaction in the antiferromagnetic compound \cite{Nejati2017} that entirely suppresses the coherent Kondo state \cite{Wetli2018}. Further, it can also be seen in Fig.~\ref{fig:Figure6}(b) that the low-frequency-averaged THz conductivity of the antiferromagnetic sample increases as the temperature is reduced. This implies that the resistivity decreases as the sample enters the antiferromagnetically-ordered phase, shown in Fig.~\ref{fig:Figure6}(c), which agrees with the electrical transport measurements \cite{Ruck1993,Loehneysen1998,Wilhelm2001}. Note that the difference in the absolute value of the resistivity measured in earlier transport measurements \cite{Ruck1993} compared to our measurements stems from different probing techniques. In the earlier transport studies, the DC values of the resistivity were measured, while in our study, we consider the frequency-averaged resistivity values at THz frequencies. 

\section{Conclusion} 
We have explored the temperature dependence of the dynamical conductivity at THz frequencies for the canonical CeCu$_{6-x}$Au$_{x}$ system. Terahertz excitation creates both intraband excitations as well as resonant interband transitions between the hybridized heavy and light parts of the conduction band. In the heavy-Fermi-liquid phase, CeCu$_6$, the intraband excitations lead to Drude response embedded within the instantaneous response. The interband transitions create a time-delayed response, featuring Kondo correlations. We have systematically separated the conductivity responses of the instantaneous signal from the delayed correlated signal. The instantaneous response of the heavy-fermion sample and the quantum-critical sample have a simple metallic Drude low-frequency behavior at high temperatures that develops into heavy-Fermi-liquid Drude and non-Fermi-liquid behavior, respectively, below the Kondo lattice temperature. In contrast, the optical conductivities obtained from the correlated, delayed response do not show Drude-like behavior at any temperatures, neither for CeCu$_6$ nor CeCu$_{5.9}$Au$_{0.1}$. We have observed a smooth transfer of the conductivity spectral weight from the CEF resonance at high temperature to the Kondo resonance at low temperature. By contrast, in CeCu$_{5.9}$Au$_{0.1}$ the spectral weight of the Kondo resonance continuously vanishes below $T_{\rm K}^*$, implying that the sample undergoes a Kondo breakdown as we approach the QCP,
and reconciling with our previous analysis in the time-domain \cite{Wetli2018, Pal2019}. This is corroborated by the $\omega/T$ scaling of the optical conductivity (real part), that we find in the scaling regime. On the other hand, the antiferromagnetic compound, CeCu$_5$Au$_1$, shows a complete absence of coherent Kondo states. Our time-resolved and phase-sensitive measurements open up the possibility to determine the optical conductivity without resorting to Kramers-Kronig analysis, and, thus, to temporally separate the purely correlated response in the time domain. This allows to distinctly identify different contributions to the dynamic conductivity. We not only reconcile the previous results obtained from infrared optical measurements, but also show a very general way to uniquely discern the dynamic processes in Kondo systems.

\section*{Acknowledgements} This work was financially supported by the Swiss National Science Foundation (SNSF) via project No. 200021{\_}178825 (M.F., S.P.), NCCR MUST via PSP 1-003448-051 (M.F., C.J.Y.), NCCR ETH FAST 3 via PSP 1-003448-054 (M.F., S.P.) and by the Deutsche Forschungsgemeinschaft (DFG) within the Cooperative Research Center SFB/TR 185 (277625399) (J.K., F.Z.). S.P. further acknowledges the support by ETH Zurich Career Seed Grant, SEED-17 18-1.

\appendix


\section{Optical conductivity analysis}

THz time-resolved spectroscopy recording the electric light field $E(t)$ is a phase-sensitive measurement which, in contrast to spectroscopy of the light intensity, gives direct access to the optical response functions of materials without resorting to Kramers-Kronig analysis. Care has been taken to ensure that the optical paths of the reference measurements and the sample measurements are identical by using two photodiodes in a crossed geometry. The crossing point corresponds to the reference (or sample) position. Any small discrepancy on the order of micrometers would reveal itself by a shift within the time traces. Such discrepancies can be adjusted later before taking the Fourier transforms of the raw data. After separating the two transients (instantaneous and delayed), care is taken once again to make sure that the total temporal window corresponding to the delayed response is the
same as that of the reference transient. The THz transients in the time-domain, i.e., $E_{\text{sample}}(t)$ and $E_{\text{refenence}}(t)$ are then converted to the frequency domain via Fourier transform. The reflectivity,
$R(\omega)$, is defined by the relation
\begin{equation}
R(\omega) = \frac{E_{\text{sample}}(\omega)}{E_{\text{reference}}(\omega)},
\label{eq_reflectivity}
\end{equation}
where $\omega = 2\pi\nu$ is the frequency, $E_{\text{sample}}(\omega)$ and
$E_{\text{reference}}(\omega)$ are the spectral distribution of the signal
reflected from the sample and from a Pt reference, respectively. The reference
consists of a 15 nm Pt film grown on a quartz substrate and placed next to the
sample inside the cryostat, so that  $E_{\text{sample}}(\omega)$ and
$E_{\text{reference}}(\omega)$ can be measured under identical experimental
conditions. The reflectivity $R_{\text{Pt}}$ of the Pt reference is
temperature independent below 100 K, so that
$E_{\text{reference}}(\omega)=R_{\text{Pt}}\,E_{\text{incident}}(\omega)$
reflects the spectrum of the incident pulse, and possible spectral features of
the Pt reference can be identified by their temperature independence, see
Sec.~II. This completes the linear response relation
(\ref{eq_reflectivity}). Our experimental geometry is set for $p$-polarized
configuration. For further details on the experimental geometry, see
supplemental material of Ref.~\cite{Pal2019}. Using Fresnel's equation, the
reflection coefficient, $r(\omega)$, for TM configuration can be obtained from the following relation:
\begin{figure}[t!]
\centering
\includegraphics[width=\columnwidth]{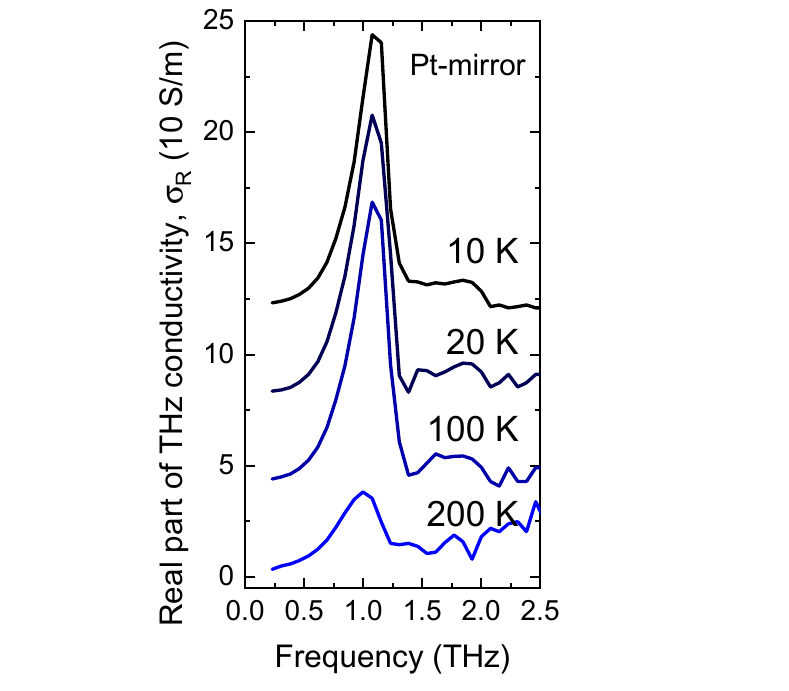}
\caption{The real part of THz optical conductivity of Pt-film (used as reference in our measurements) for a set of temperatures. The curves are offset for clarity.}
\label{fig:reference}
\end{figure}
\begin{equation}
r(\omega) = \frac{n_{1}cos(\theta_{2}) - n_{2}cos(\theta_{1})}{n_{1}cos(\theta_{2}) + n_{2}cos(\theta_{1})}, 
\label{eq_fresnel}
\end{equation}
where $n_{1}$ and $n_{2}$ are the refractive indices of air and the sample,
respectively. $\theta_{1}$ and $\theta_{2}$ are the angle of incidence and the
angle of refraction, respectively. In our experimental setup, the incidence
angle $\theta_{1}$ is 45$^{\circ}$ and the refractive index of air is $n_{1} =
1$. Using  Snell's law ($n_{1}sin\theta_{1}=n_{2}sin\theta_{2}$),
Eq.(\ref{eq_fresnel}) can be expressed as
\begin{equation}
r(\omega) = \frac{-n_{2}^{2}(\omega)cos(45^{\circ}) + \sqrt{n_{2}^{2}(\omega) - sin^{2}(45^{\circ})}}{n_{2}^{2}(\omega)cos(45^{\circ}) + \sqrt{n_{2}^{2}(\omega) - sin^{2}(45^{\circ})}}.
\label{eq_transfer}
\end{equation}
This further simplifies to
\begin{equation}
r(\omega) = \frac{-n_{2}^{2}(\omega) + \sqrt{2n_{2}^{2}(\omega) - 1}}{n_{2}^{2}(\omega) + \sqrt{2n_{2}^{2}(\omega) - 1}},
\label{eq_n}
\end{equation}
where the refractive index of the sample, $n_{2}(\omega) = n'(\omega) + i\,n''(\omega)$ is the only variable. It is to be noted that $n_{2}$ is a complex, frequency-dependent number. The complex refractive index can then be
obtained from the experimentallly determined $R(\omega)$ at various
temperatures. Using the Drude-Lorentz model, we then obtain the complex THz
optical conductivity. To extract out the conductivity response from the
delayed pulse, we cut the original data in two parts, namely the instantaneous
and the delayed responses. The choice of the position that separates the
instantaneous response and the delayed response is purely based on the
envelope function used to model the delayed response see supplemental material of Ref.~\cite{Wetli2018}. We cut the data such that the envelope function has zero weight and we include as much of the data as possible from the wings of the instantaneous response. We kept this same, i.e. 3.5 ps, as in our previous works \cite{Wetli2018,Pal2019}. A slight change of this boundary, taking care not to cut the data in the delayed response or the instantaneous response with the wiggles, has no effect in our analysis. Frequencies lower than 0.2\,THz are more sensitive to such changes mainly due to lower spectral weight in this region and hence we did not consider it in our discussions.

\begin{figure}[t!]
\centering
\includegraphics[width=\columnwidth]{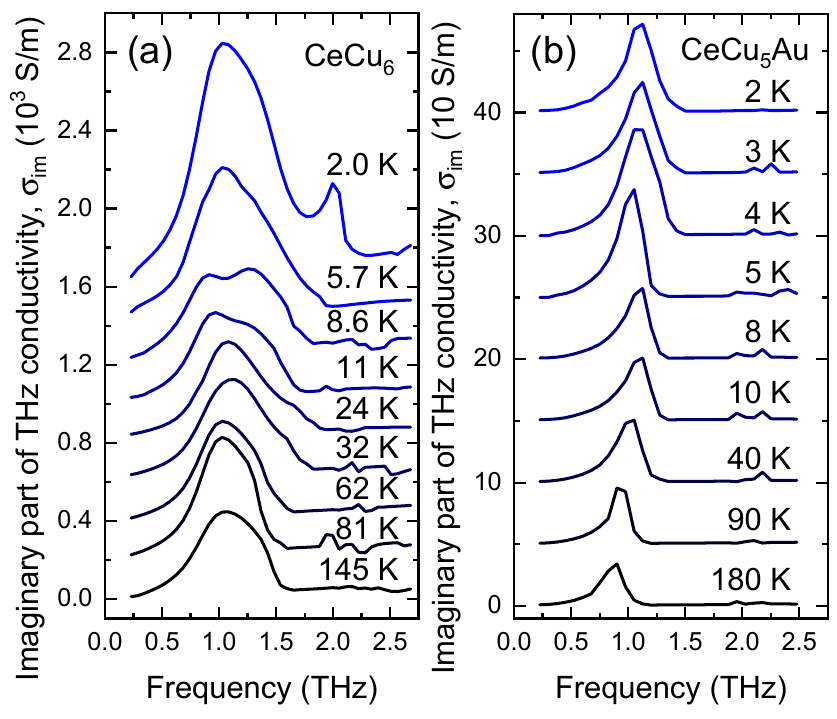}
\caption{The imaginary part of the THz optical conductivity of (a) CeCu$_{6}$ and (b) CeCu$_{5}$Au$_{1}$ samples obtained from the instantaneous response as a function of the sample temperature.}
\label{fig:imag_1}
\end{figure}

\section{Conductivity peak at 1 terahertz}

Even the Pt film is not an ideal mirror, but has some frequency-dependent response. For the analysis in the time domain of our earlier works \cite{Pal2019,Wetli2018}, this was not an issue. In the frequency-domain analysis, it leads to a nearly temperature-independent  conductivity peak near 1\,THz. In order to identify this feature, we carried out the conductivity analysis coming from the instantaneous response of the Pt film. For this analysis we used the bare THz pulse as a reference. Fig.\ref{fig:reference} shows the real part of the THz conductivity of the Pt film for various temperatures between 200 and 10~K. We find that the peak at approx. 1\,THz is temperature-independent below 100~K with no additional features that we otherwise obtain from the CeCu$_{6-x}$Au$_{x}$ samples. The temperature-independent feature visible in the instantaneous spectra of Figs.~3 and 4 of the main paper is, thus, due to the Pt mirror, and should not be taken into account in the analysis of the
CeCu$_{6-x}$Au$_{x}$ response.

\begin{figure}[b!]
\centering
\includegraphics[width=\columnwidth]{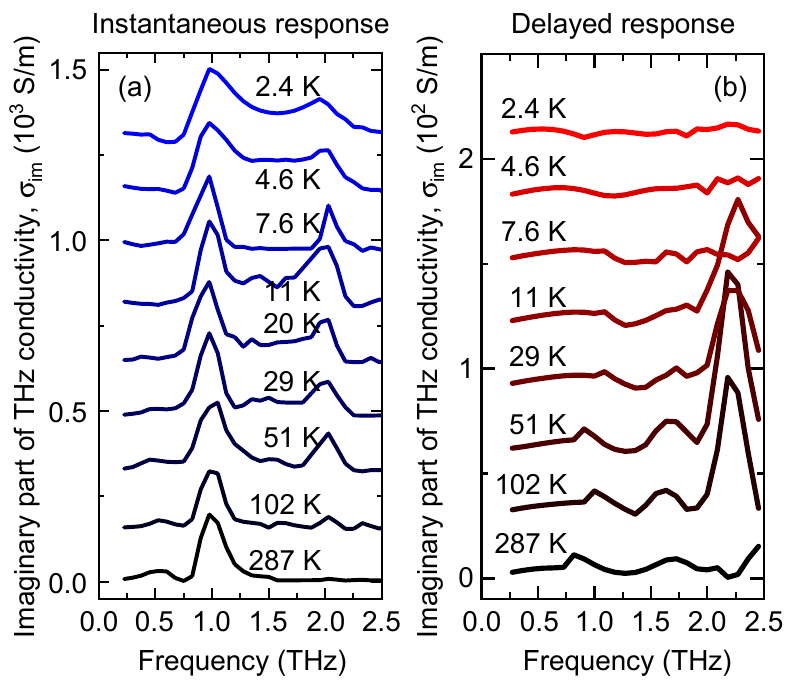}
\caption{The imaginary part of the THz optical conductivity of CeCu$_{5.9}$Au$_{0.1}$ obtained from (a) the  instantaneous response and (b) the delayed response as a function of sample temperature.}
\label{fig:imag_2}
\end{figure}

\section{Imaginary part of optical conductivity}

Figs.~\ref{fig:imag_1}(a) and \ref{fig:imag_1}(b) show the imaginary components of the THz optical conductivity obtained from the instantaneous response of the heavy-fermion compound, CeCu$_{6}$ and the antiferromagnetic
compound, CeCu$_{5}$Au, respectively, at several temperatures. The imaginary components of the THz optical conductivity from the instantaneous and delayed response of the quantum critical compound, CeCu$_{5.9}$Au$_{0.1}$ are shown in Figs.~\ref{fig:imag_2}(a) and \ref{fig:imag_2}(b), respectively.

\section{Critical scaling analysis}
We performed a critical scaling analysis of the THz optical conductivity \cite{Prochaska2020}, obtained from the instantaneous response, for the quantum-critical sample, CeCu$_{5.9}$Au$_{0.1}$, and for temperatures below the Kondo temperature of $T_{K}^{*}\approx 8$\,K (c.f.~Sec.~V), i.e., in the region of the quantum-critical fan.
First, we find that in our sample a possible, noncritical background conductivity is small enough compared to the diverging, critical part, so that background subtraction in the resistivity \cite{Prochaska2020} is not necessary in order to obtain scaling of the real part of the conductivity (see below). For our experimental setup, it would be difficult to determine the background, because the size of a noncritical Drude response cannot be pinned to the DC conductivity. 

For the scaling analysis of the real part, we convert the THz frequencies
$\nu=\omega/2\pi$ to the dimensionless parameter $x=\hbar\omega/k_{B}T$ and
plot, for all frequencies $\nu$ ranging from 0.27 to 1\,THz and for all
temperatures below 8~K (Kondo temperature $T_K^*$) the function $y^{data}(x) =
\text{Re}[\sigma_{\text{inst.}}(\omega)]\,T^{\alpha}$, where $\alpha$ is still
a free parameter of order 1. We then fit a power law, $y^{fit}(x)=A*x^{s}$, to
all the data points $y^{data}(x)$, with fit parameters $s$ and $A$. We repeat
this procedure for different values of $\alpha$ and compute each time the
relative standard deviation of the fit,
\begin{equation}
\chi = \sqrt{\frac{1}{N} \sum_{\omega_i,T_i}\frac{(y^{\rm data}_i-y_i^{\rm fit})^2} {(y_i^{\rm data})^2}},
\end{equation}
where the sum runs over all data points $(\omega_i,T_i)$ in the scaling regime. The resulting $\chi$ is shown in the inset of Fig.~5(c) of the main paper as a function of $\alpha$. The scaling exponent $\alpha$ and corresponding value of $s$ are determined as those values for which $\chi$ obtains a minimum, i.e. for $\alpha\approx 1.0$ and $s\approx 1.3$ (see Fig.~5(c)).

For a scaling analysis of the imaginary part of the optical conductivity, we find that background subtraction would be needed, since the imaginary part of a Drude background has strong frequency dependence with a sign
change at $\omega =0$, according to the Kramers-Kronig relation. We, therefore, did not perform a scaling analysis of the imaginary part, as this is not the focus of the present paper.

\begin{figure}[t!]
\centering
\includegraphics[width=\columnwidth]{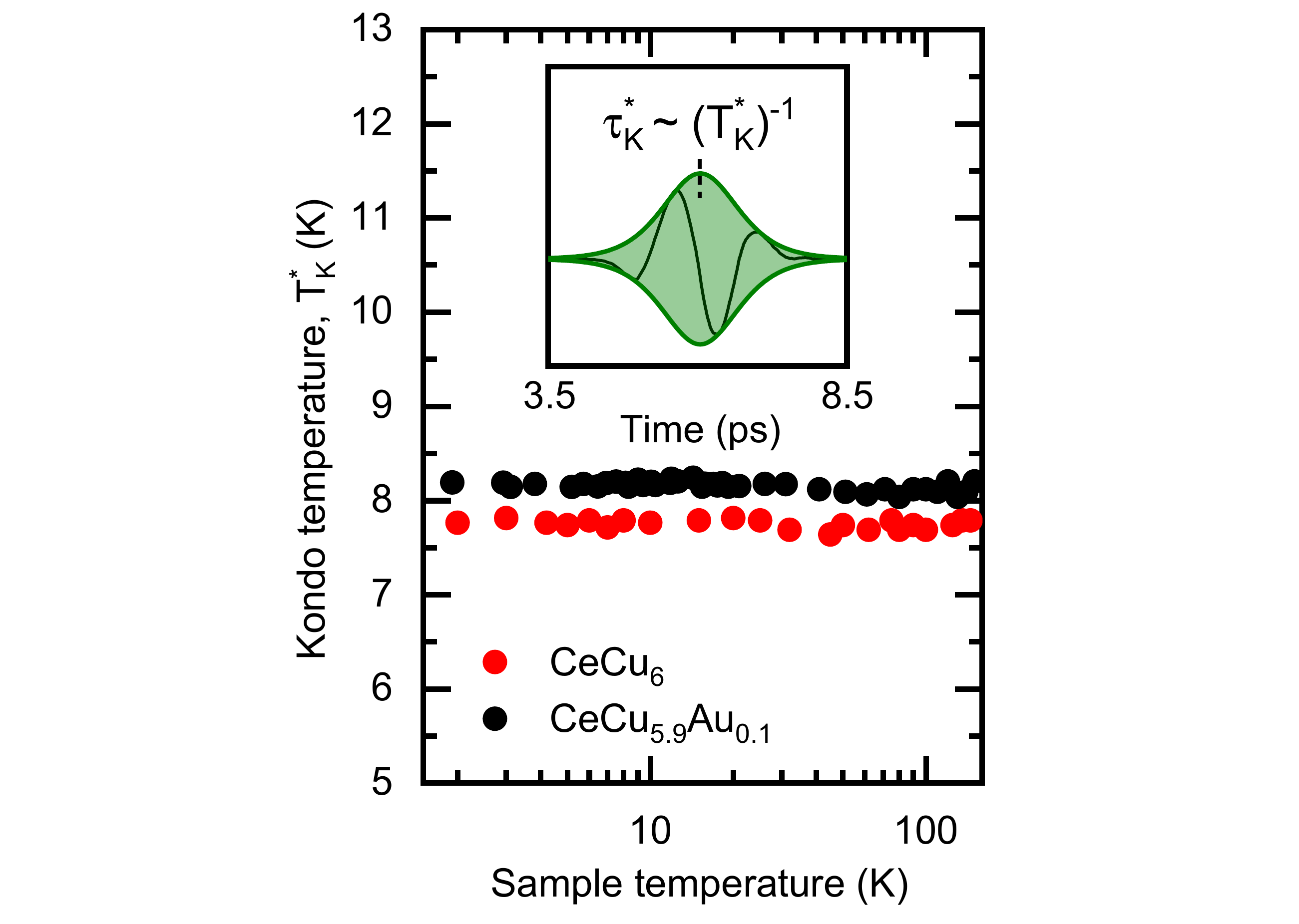}
\caption{The evolution of the lattice Kondo temperature $T_K^*$ as a function of sample temperature. $T_K^*$ is obtained from the position of the envelope of the delayed pulse as sketched in the inset. Throughout the temperature range, the Kondo temperature remains constant.}
\label{fig:dep_Kondo}
\end{figure}

\section{Temperature dependence of lattice Kondo temperature} 
In our earlier investigation \cite{Wetli2018}, we showed that the lattice Kondo temperature, extracted from the delay time $\tau_{K}^{*}$ as $T_{K}^{*}=\hbar/k_B\tau_{K}^{*}$, remains constant at low temperatures up to 25\,K. Here, we expand the temperature range further up to 150\,K, see Fig.~\ref{fig:dep_Kondo}. Within the experimental resolution, $T_{K}^{*}$ stays at the same value, indicating that the energy scale for the formation of
Kondo quasiparticles remains constant.

\end{document}